\documentclass[lettersize,journal]{IEEEtran}
\usepackage{amsmath,amsfonts}
\usepackage{algorithmic}
\usepackage{algorithm}
\usepackage{array}
\usepackage[caption=false,font=normalsize,labelfont=sf,textfont=sf]{subfig}
\usepackage{textcomp}
\usepackage{stfloats}
\usepackage{url}
\usepackage{verbatim}
\usepackage{graphicx}
\usepackage{cite}
\begin{document}

\title{Integrated balanced homodyne detector using CMOS capacitive-feedback TIA for quantum measurements}

\author{Sarah Bastiaens, Cédric Bruynsteen, Simone Cammarata, Axl Bomhals, Leandro da Silva, Michiel Van Osta, Johan Bauwelinck, and Xin Yin%
\thanks{This work was supported by European Union’s Horizon Europe research and innovation program under the project “Quantum Security Networks Partnership” (QSNP, 101114043); and FWO-Weave grant G092922N.}
\thanks{All authors are with IDLab, INTEC, imec-Ghent University, 9052 Ghent, Belgium (mail: sarah.bastiaens@ugent.be)}
\thanks{Cédric Bruynsteen now with Nvidia, 9050 Ghent, Belgium}}

\markboth{Journal of Lightwave Technology \LaTeX\ Class Files,~Vol.~, No.~, May~2026}%
{Sarah Bastiaens, Cédric Bruynsteen \MakeLowercase{\textit{et al.}}:Integrated balanced homodyne detector using CMOS capacitive-feedback TIA for quantum measurements}


\maketitle

\begin{abstract}
A low‑noise, high‑speed balanced homodyne detector is demonstrated, combining a capacitive‑feedback transimpedance amplifier implemented in 28 nm CMOS with a custom photonic integrated circuit fabricated in imec’s iSiPP200 silicon photonics platform. The receiver achieves a 3‑dB bandwidth of 7.6 GHz, a maximum shot noise clearance of 27 dB, a shot noise‑limited bandwidth of 13.7 GHz and a common‑mode rejection ratio up to 54.8 dB. These results demonstrate the suitability of CMOS capacitive‑feedback transimpedance amplifiers for demanding quantum and coherent sensing optical applications, including continuous‑variable quantum key distribution, quantum random number generation, and optical coherence tomography. 
\end{abstract}


\section{Introduction}

The ability to detect and process extremely weak optical signals is fundamental to a wide range of sensitive optical applications, such as quantum key distribution (QKD) \cite{QKD,cvqkd,tia_noise}, quantum random number generation (QRNG) \cite{Axl,c_qrng,qrng_c}, quantum computing \cite{qc}, the characterization of quantum states \cite{q_st,st_q}, coherent lidar \cite{li,lid}, coherence tomography \cite{ct,oct}, and gas sensing \cite{ga,gs}. Achieving the quantum‑noise limit in these applications is challenging with traditional direct‑detection receivers, which are fundamentally constrained by laser intensity noise and detector noise. Balanced homodyne detection (BHD) provides a well‑established solution, by mixing the incoming signal with a strong local oscillator (LO) in a beam splitter and subtracting the photocurrents of two matched photodiodes (PDs), BHDs simultaneously suppress common‑mode noise and enable the measurement of optical field quadratures.

Transimpedance amplifiers (TIAs) are critical components of BHDs, because they convert a weak photocurrent into a measurable voltage. The TIA has a significant impact on the overall bandwidth, noise performance, and linearity of the receiver. In practice, the TIA often accounts for the dominant noise source in BHD systems \cite{tia_noise}.

Traditional BHDs have historically relied on discrete, off‑the‑shelf PDs and TIAs. However, as state‑of‑the‑art systems push toward multi‑GHz bandwidths, these discrete architectures are increasingly constrained by packaging parasitics that degrade both bandwidth and noise performance \cite{cedric}. Modern integrated BHD implementations address this limitation by co‑designing a TIA directly with the photonic integrated circuit (PIC). This co‑integration minimizes parasitics and significantly improves the noise performance, bandwidth, and tighter overall system integration compared to traditional discrete solutions.

A clear trend in the current state‑of‑the‑art technology is the monolithic integration of photonic and electronic components, which has advanced rapidly in recent years \cite{Mono_sim,bicmos}. These platforms offer extremely compact receiver footprints and open the door for scalable wafer‑level quantum photonic systems. Despite this progress, a persistent performance bottleneck lies in the quantum efficiency of the PDs available in these technologies, which typically remains limited to 40–60\%. This constraint directly impacts the shot‑noise clearance and limits the achievable sensitivity of fully integrated receivers.

Recent demonstrations have illustrated both the strengths and limitations of today’s most advanced technologies. For example, a high‑performance TIA implemented in 100 nm GaAs pHEMT technology \cite{cedric} achieves a 3 dB bandwidth of 1.5 GHz and a maximum shot‑noise clearance of 28 dB, which remains competitive with leading BHD designs. However, GaAs‑based approaches face challenges that limit their suitability for next‑generation integrated receivers, including high power consumption, lower transistor‑scaling potential, reduced integration density, and substantially higher fabrication cost relative to silicon‑based platforms.

In contrast, advanced complementary metal-oxide semiconductor (CMOS) technology offers a compelling path. Deep‑submicron CMOS provides a high transistor density, low power consumption, and the possibility of co‑integrating analog, RF, and digital signal‑processing circuitry on the same die. Moreover, co‑integration with silicon photonics eliminates many packaging parasitics that traditionally limit bandwidth and noise, enabling performance levels that are difficult to achieve with discrete components.

In this work, we present a capacitive‑feedback TIA implemented in 28 nm CMOS, interfaced with a custom silicon photonic chip fabricated using imec’s iSiPP200 platform, creating a fully integrated, high‑performance phase-diverse BHD. We report detailed electrical and opto‑electrical characterization of the system and demonstrate the potential of advanced CMOS to achieve state‑of‑the‑art performance in compact, scalable, low power receivers.

\section{Design of the Balanced Homodyne Detector}

\begin{figure*}[!t]
\centerline{\includegraphics[width=0.7\textwidth]{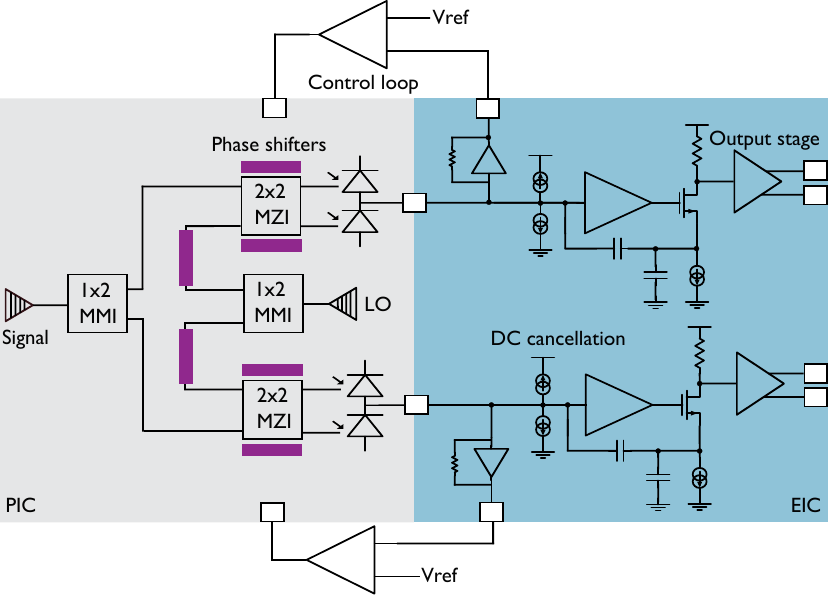}}
\caption{Simplified schematic of the designed BHD, comprising of a PIC, a TIA and an external feedback loop.}
\label{BHD_shem}
\end{figure*}

A schematic overview of the designed BHD is shown in Figure \ref{BHD_shem}. The system consists of three distinct components: a PIC, an electronic integrated circuit (EIC) containing the TIA, and an external feedback loop used for DC balancing.

\subsection{Photonic Integrated Chip}

The PIC is implemented using the iSiPP200 process from imec's silicon photonics platform. The optical signals from the fibers are coupled into the PIC through grating couplers. It has two channels that enable measurement of both the in‑phase (I) and quadrature (Q) components of the optical field. Channel separation is achieved using a 1×2 multi‑mode interferometer (MMI). One input carries a high power local oscillator (LO), whereas the other carries the input signal.

Within each channel, the signal and the LO are combined in a tunable Mach–Zehnder interferometer (MZI). The power‑splitting ratio of the MZI is controlled by integrated thermal phase shifters, which are actuated by applying a (biasing) voltage to the heater bond pads through the wirebonds. The resulting optical fields are then directed to a pair of balanced PIN PDs. The PDs exhibit a nominal responsivity of 1.1 A/W at a wavelength of 1550 nm and a 6‑dB opto‑electrical bandwidth of approximately 11.2 GHz under a 1 V reverse bias. The PDs feature a low junction capacitance, minimizing
the capacitive loading at the TIA input. The photocurrents from the two PDs are subtracted at a common node and delivered directly to the TIA input via short wirebonds to limit parasitic inductance.

Owing to fabrication tolerances, the responsivities of the two PDs are not perfectly matched, resulting in a residual DC current at the TIA input. This mismatch can shift the DC operating point of the TIA and degrade its performance. Additionally, imperfect balancing prevents full suppression of LO relative intensity noise (RIN). This residual DC current can be mitigated by adjusting the optical power delivered to each PD through fine-tuning the MZI heaters. 

\subsection{Transimpedance Amplifier}

The TIA converts the small input current generated by the balanced PDs into a voltage suitable for subsequent processing by an analog-to-digital converter (ADC), while introducing minimal electronic noise. The TIA is implemented in 28 nm CMOS technology. CMOS technology offers several advantages over III–V technologies, including lower fabrication cost, reduced power consumption, high integration density, and compatibility with large‑scale manufacturing. This advanced technology node provides a high transition frequency $f_t$ above 300GHz, enabling large achievable bandwidths. Its primary drawback is inferior noise performance owing to lower carrier mobility and the presence of short-channel effects in deeply scaled CMOS nodes, which introduces excess noise \cite{Exnoise}.

To mitigate these limitations, a capacitive-feedback topology is adopted. Replacing the feedback resistor with a capacitor eliminates the dominant resistor thermal noise present in shunt‑feedback TIAs, improving the noise performance in the frequency range of several GHz, which is especially relevant for BHDs for quantum applications. The proposed topology is analyzed in detail and compared with the shunt‑feedback alternative, followed by a description of the complete TIA design.

\subsubsection{Capacitive-Feedback Topology}

\begin{figure}[!t]
\centering
\includegraphics[width=3in]{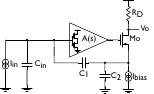}
\caption{Simplified schematic of a capacitive-feedback TIA.}
\label{Cap-fb}
\end{figure}

Figure \ref{Cap-fb} shows the capacitive feedback topology. Its transfer function is given by

\begin{equation}
Z_T(s) = \frac{-g_m R_D(C_1 (A(s)+1) + A(s) C_2)}{g_m(C_1 (A(s)+1) + C_{in}) + s(C_1 C_2 + C_1 C_{in} + C_2 C_{in})}
\label{eq:TF}
\end{equation}

where $C_{in}$ is the total parasitic input capacitance (provided by the photodiodes, TIA input, bondpad, and ESD protection) and $g_m$ is the transconductance of the output transistor $M_o$. Assuming a single-pole amplifier A(s) model with a high DC gain $A_0$, the DC transimpedance can be simplified to:

\begin{equation}
DC_{gain} = - R_D (1+\frac{C_2}{C_1})
\end{equation}

Assuming that the transfer function (Eq. \ref{eq:TF}) follows a Butterworth response, $A_0 \gg 1 $, and $C_2 \gg C_1 $, the 3‑dB bandwidth can be approximated as

\begin{equation}
\omega_{3dB} = \frac{g_m C_1 \sqrt{2 A_0 (A_0+1)}}{C_2 (C_1 + C_{in})}
\end{equation}

Thus, the transimpedance gain is determined by the resistor $R_D$ and the capacitor ratio rather than the absolute resistor value alone, making the topology more tolerant to process, voltage, and temperature (PVT) variations than shunt‑feedback TIAs. The DC transimpedance gain is also independent of the transconductance $g_m$, thereby decoupling the DC gain from the TIA bandwidth. In shunt-feedback, this degree of freedom is not present, because the bandwidth and DC gain are both
set by the feedback resistance \cite{cedric}.

The noise performance is another key design consideration. For the capacitive-feedback topology, the total input-referred noise is dominated by two primary contributions. At lower frequencies, the thermal noise generated by the biasing of the output transistor $M_o$, denoted $I_{n,bias}$ is dominant. At higher frequencies, the thermal noise of the input transistors of amplifier A(s), represented by $V_{r,in}$, is the primary contributor. All the remaining noise sources are calculated to be negligible relative to the two dominant mechanisms.

The input-referred current noise contribution of the bias transistor is

\begin{equation}
I_{n,bias}^2  = \Gamma 4 k T g_{m,bias} \Big( \frac{(A(s)+1)C_1+ C_{in}}{(A(s)+1)C_1+ A(s) C_2}\Big)^2
\end{equation}

where k is Boltzmann constant, T is the absolute temperature, $\Gamma$ is Ogawa’s excess noise factor, and $g_{m,bias}$ is the transconductance of the bias transistor. Assuming $C_2 \gg C_1 $ and $A_0 \gg 1$, then for low frequencies the equation can be simplified to:

\begin{equation}
I^2_{n,bias} = \Gamma 4 k T g_{m,bias}\Big(\frac{A_0 C_1 + C_{in}}{A_0 C_2}\Big)^2 
\end{equation}

The input-referred current noise contribution of the input transistors $V_{r,in}$ in $A(s)$ is

\begin{equation}
I^2_{n,in} = 2 \frac{\Gamma 4 k T}{g_{m,in}} \Big(\frac{ A(s) s (C_1 C_2 + C_1 C_{in} + C_2 C_{in})}{ (A(s)+1) C_1 + A(s) C_2}\Big)^2
\end{equation}

where $g_{m,in}$ is the average transconductance of the NMOS and PMOS devices in the input inverter. Under the same assumptions, this simplifies to:

\begin{equation}
I^2_{n,in} = 2 \frac{\Gamma 4 k T}{g_{m,in}} s^2(C_1 + C_{in})^2
\end{equation}

The total input‑referred current noise is therefore

\begin{equation}
I^2_{n,in} = 2 \frac{\Gamma 4 k T}{g_{m,in}} s^2(C_1 + C_{in})^2 + \Gamma 4 k T g_{m,bias}\Big(\frac{A_0 C_1+C_{in}}{A_0 C_2}\Big)^2 
\end{equation}

\begin{figure}[!t]
\centerline{\includegraphics[width=0.6\textwidth]{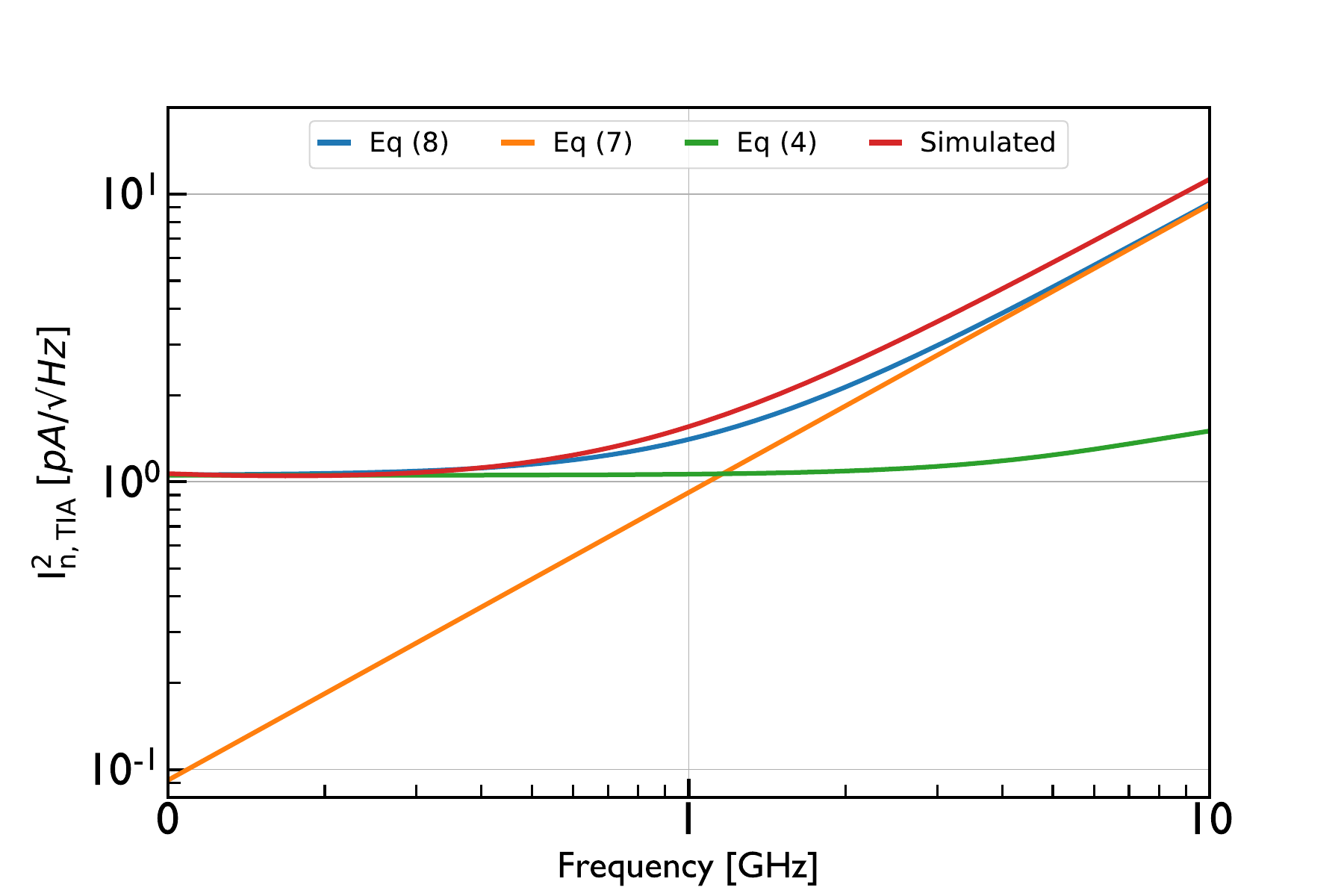}}
\caption{Analytical noise model of a capacitive-feedback TIA with comparison to circuit-level simulations of the complete circuit (Cadence virtuoso).}
\label{noise}
\end{figure}

Figure \ref{noise} compares the analytical noise model with the numerical simulations of the complete circuit in Cadence Virtuoso. The analytical model closely matches the simulated behavior. The difference is caused by simplification of the analytical model.

\subsubsection{Comparison of Shunt-Feedback and Capacitive-Feedback Topologies}

\begin{figure*}[!t]
\centering
\subfloat[]{\includegraphics[width=3.5in]{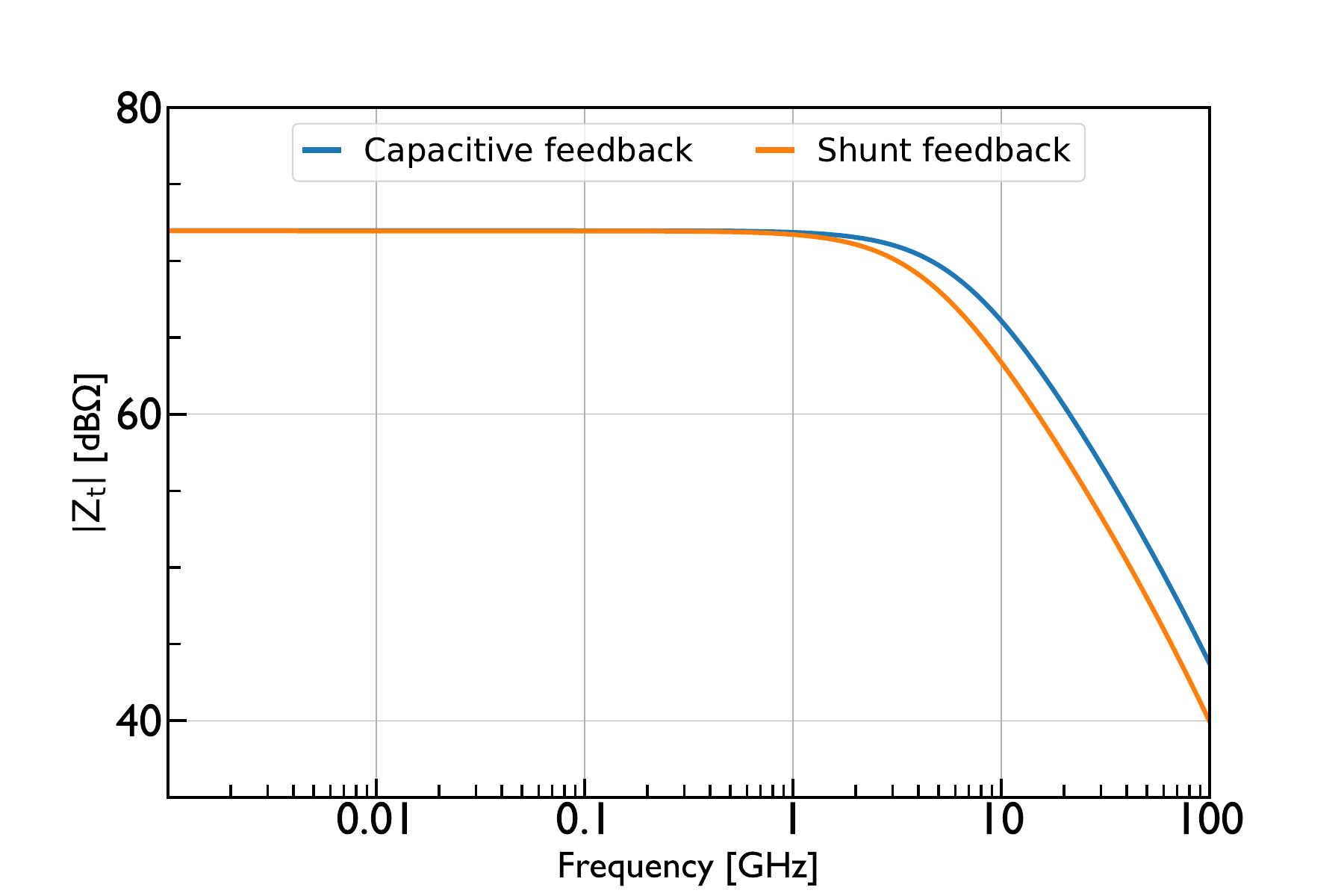}%
\label{TF_comp}}
\hfil
\subfloat[]{\includegraphics[width=3.5in]{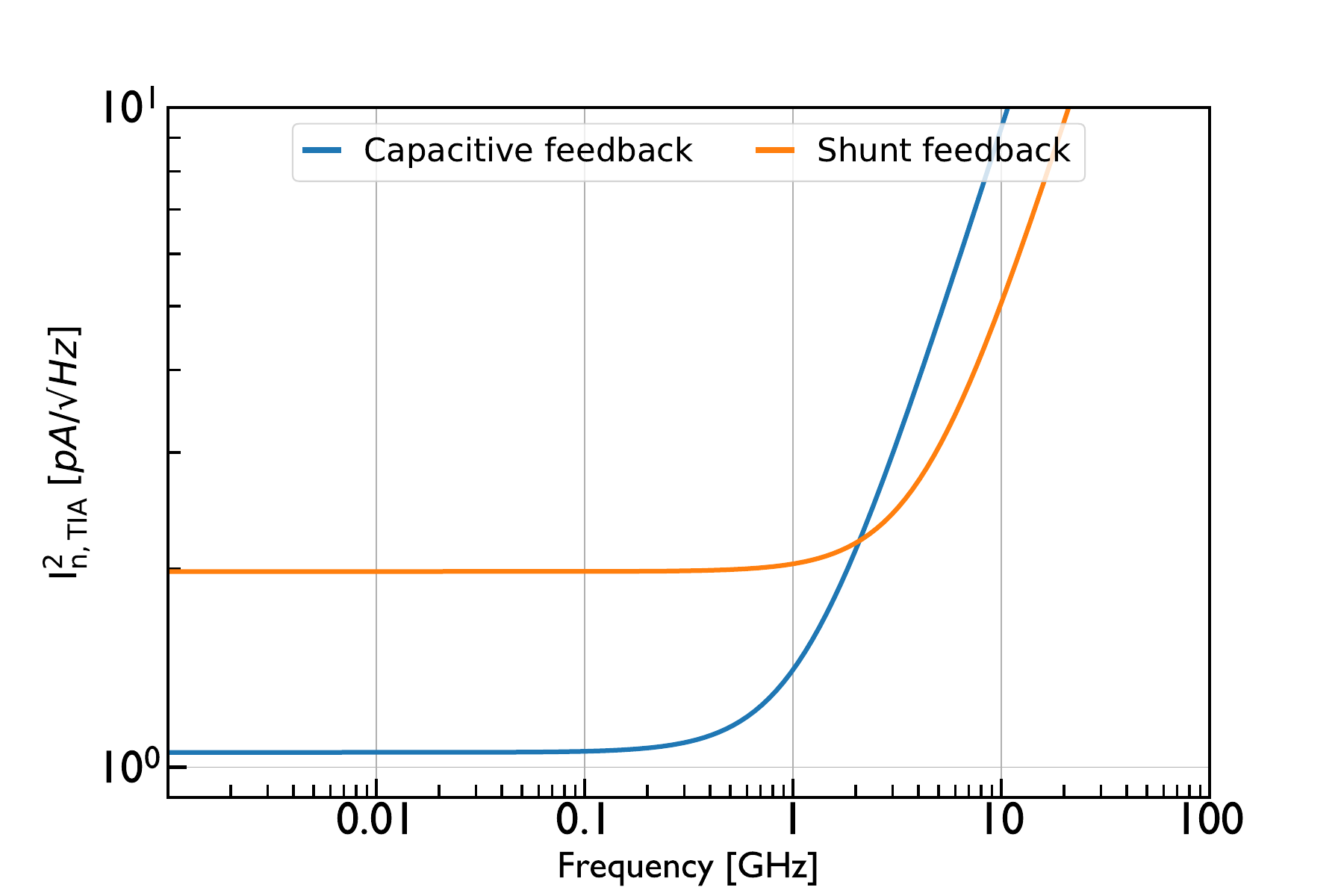}%
\label{noise_comp}}
\caption{(a) Comparison of the transfer functions for capacitive-feedback and shunt-feedback TIAs, configured for the same DC gain of 72 dB$\Omega$ (b) Noise performance comparison between capacitive-feedback and shunt-feedback TIAs, both configured for a DC transimpedance gain of 72 dB $\Omega$}
\end{figure*}

Figure \ref{TF_comp} compares the transfer functions of the capacitive-feedback and shunt-feedback topologies for an identical DC gain of 72 dB$\Omega$. In both cases, an analytical model was used, with a single-pole amplifier A(s) model. Under this constraint, the capacitive-feedback topology achieves a slightly higher bandwidth.

Figure \ref{noise_comp} shows a comparison of the input‑referred noise of both topologies, which is also evaluated at a DC gain of 72 dB$\Omega$. At low frequencies, the capacitive-feedback TIA exhibits lower noise, whereas at high frequencies, its noise increases more rapidly. Consequently, below this corner frequency, the capacitive-feedback topology provides superior noise performance, whereas above this frequency, the shunt‑feedback topology is advantageous.

\subsubsection{Design of the Transimpedance Amplifier}

\begin{figure*}[!t]
\centering
\includegraphics[width=6in]{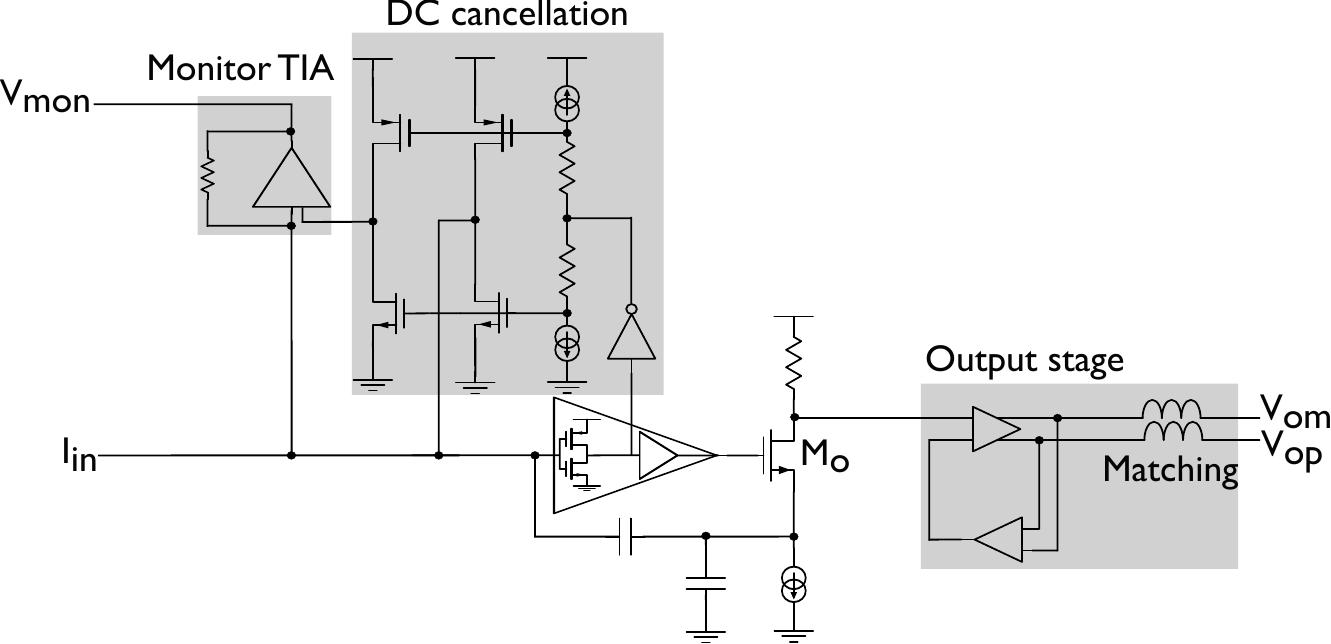}
\caption{Schematic of the proposed TIA.}
\label{TIA_deep}
\end{figure*}

Figure \ref{TIA_deep} shows a detailed schematic of the proposed TIA. The front end employs an inverter‑based amplifier. In advanced CMOS nodes, inverter‑based gain stages provide a high bandwidth because the effective transconductance is the sum of the NMOS and PMOS $g_m$ values \cite{inv}. They also operate efficiently under a limited voltage headroom, which is a common constraint in deeply scaled technology. Their main limitation is their reduced power‑supply rejection ratio (PSRR).

Following the output transistor $M_o$, an output stage is incorporated consisting of an output driver, a single-ended-to-differential converter, and an output matching network designed to achieve a 50 $\Omega$ output impedance. Tunability is integrated into the TIA to ensure robust performance across process variations, controlled through an on-chip digital core.

A significant limitation of the capacitive‑feedback topology is its sensitivity to DC input currents. Due to the integrator‑like behavior of the feedback capacitor, even small DC currents can drive the front end to saturation. To mitigate this, an on‑chip DC‑cancellation circuit is added at the input of the EIC. The DC input current is sensed at the front end of the TIA in the DC cancellation circuit using a crossover-distortion-based detector. The output of this detector controls the NMOS and PMOS current sources that implement the DC-cancellation mechanism. In the absence of DC input current, the output of the sensing circuit is at mid-supply, and the current sources are in the sub-threshold regime. When a DC current is present, the detector output shifts away from the mid-supply, and the direction of this shift determines whether the NMOS or PMOS branch sinks or sources the current. Tunability is incorporated to ensure robust operation, because the crossover point is sensitive to process variations. Additionally, tunability is provided by adjusting the number of active devices in the cancellation circuit, allowing more current to be compensated. 

In addition to compensating for the input DC current, the cancellation loop also establishes the DC bias point of the inverter-based amplifier, by injecting current directly at the TIA input, the loop forces the inverter stage to settle at a stable voltage. This feedback loop also suppresses the low-frequency components, and thus defines the low‑frequency cutoff of the receiver. 

However, introducing cancellation circuitry incurs a noise penalty: the noise contributed by the DC‑cancel transistors appears directly at the TIA input, degrading the overall sensitivity. To address this limitation, a second mechanism is employed: an external feedback loop that compensates for DC imbalance by leveraging the heaters on the PIC.

\subsubsection{Design of the External Feedback Loop}

\begin{figure*}[!t]
\centering
\subfloat[]{\includegraphics[width=3.5in]{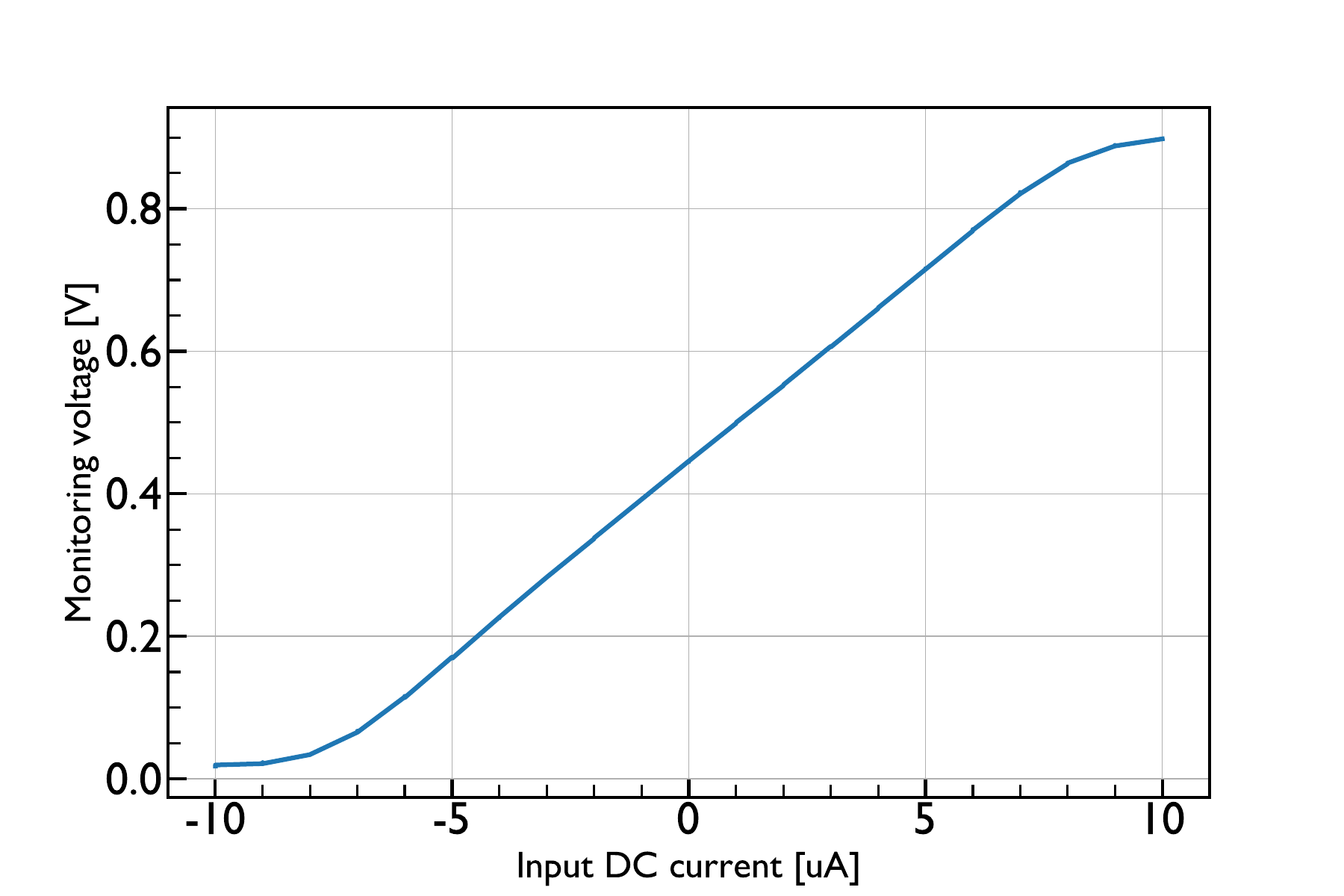}%
\label{Vmon}}
\hfil
\subfloat[]{\includegraphics[width=3.5in]{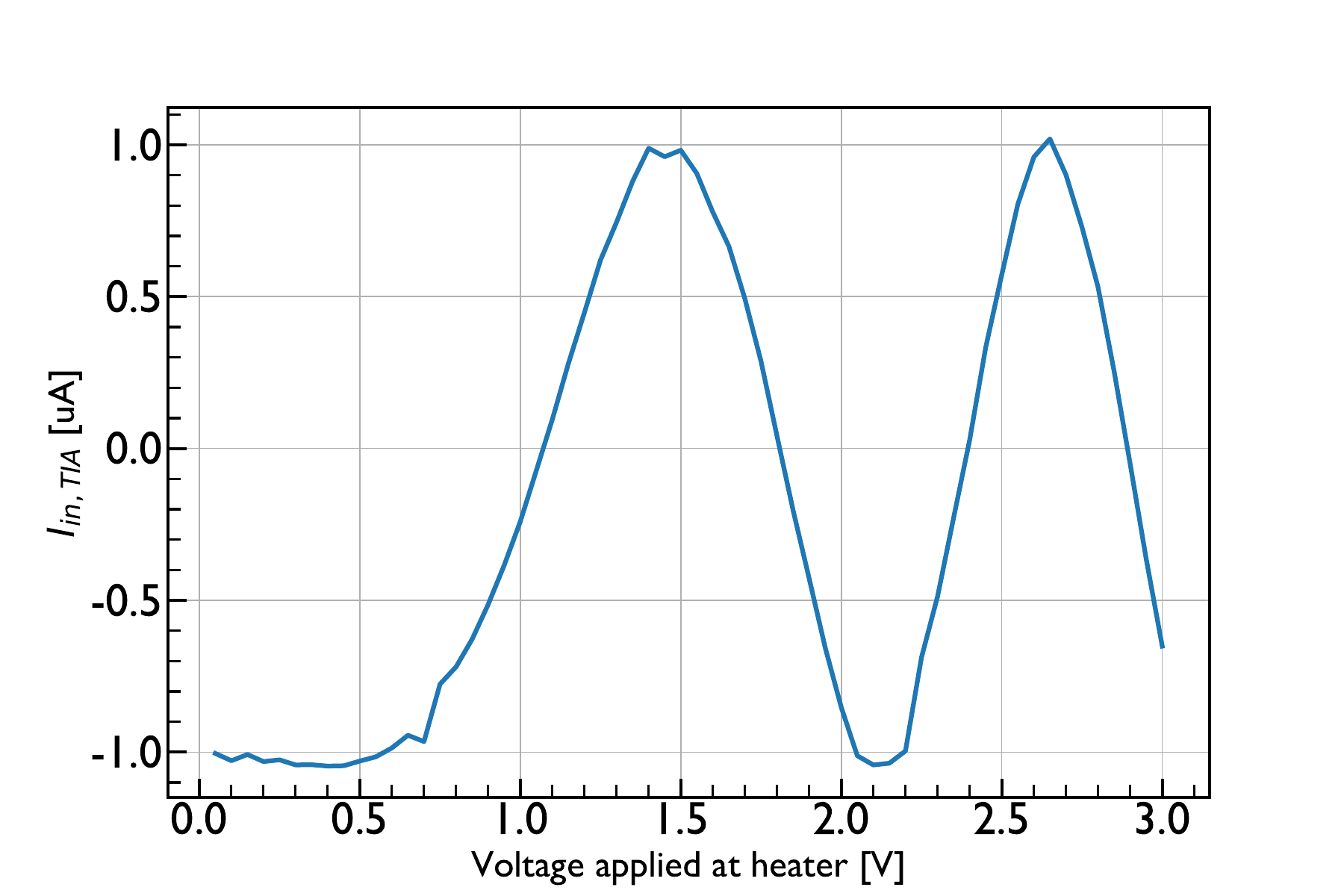}%
\label{VMZI}}
\caption{(a) Output voltage of the on-chip auxiliary TIA as function of the applied DC input current. (b) Difference in photocurrent between the balanced PDs as function of the voltage applied to the MZI heaters.}
\label{fbl}
\end{figure*}

The DC current entering the EIC is monitored using an auxiliary on-chip monitor TIA, whose output is routed to a dedicated bond pad. This signal is used in an external control loop that drives the heater of the tunable MZI on the PIC, thereby adjusting the optical power-splitting ratio. Figure \ref{Vmon} illustrates the monitoring output as a function of the DC current, depending on the current polarity, the output voltage increases or decreases accordingly. The DC current at the TIA input is effectively minimized by rebalancing the optical power incident on the balanced PDs. The external loop may be implemented either digitally or in an analog form (lower complexity as no ADC, DAC, etc. are required). 

 In analog implementation, the control loop receives two inputs: a reference voltage and the output of the monitoring TIA. An operational amplifier amplifies the difference between these voltages, generating a control voltage that drives the heater of the tunable MZI. Figure \ref{VMZI} illustrates the periodic behavior of the MZI as a function of the applied heater voltage. The MZI characteristic has multiple points where the net current is zero. Therefore, a voltage divider is used to restrict the voltage applied to the heater to an interval around the first net zero point. A buffer stage is included to drive the low-resistance heater element.

\section{Characterization of the Balanced Homodyne Receiver}

\begin{figure}[!t]
\centering
\includegraphics[width=2.5in]{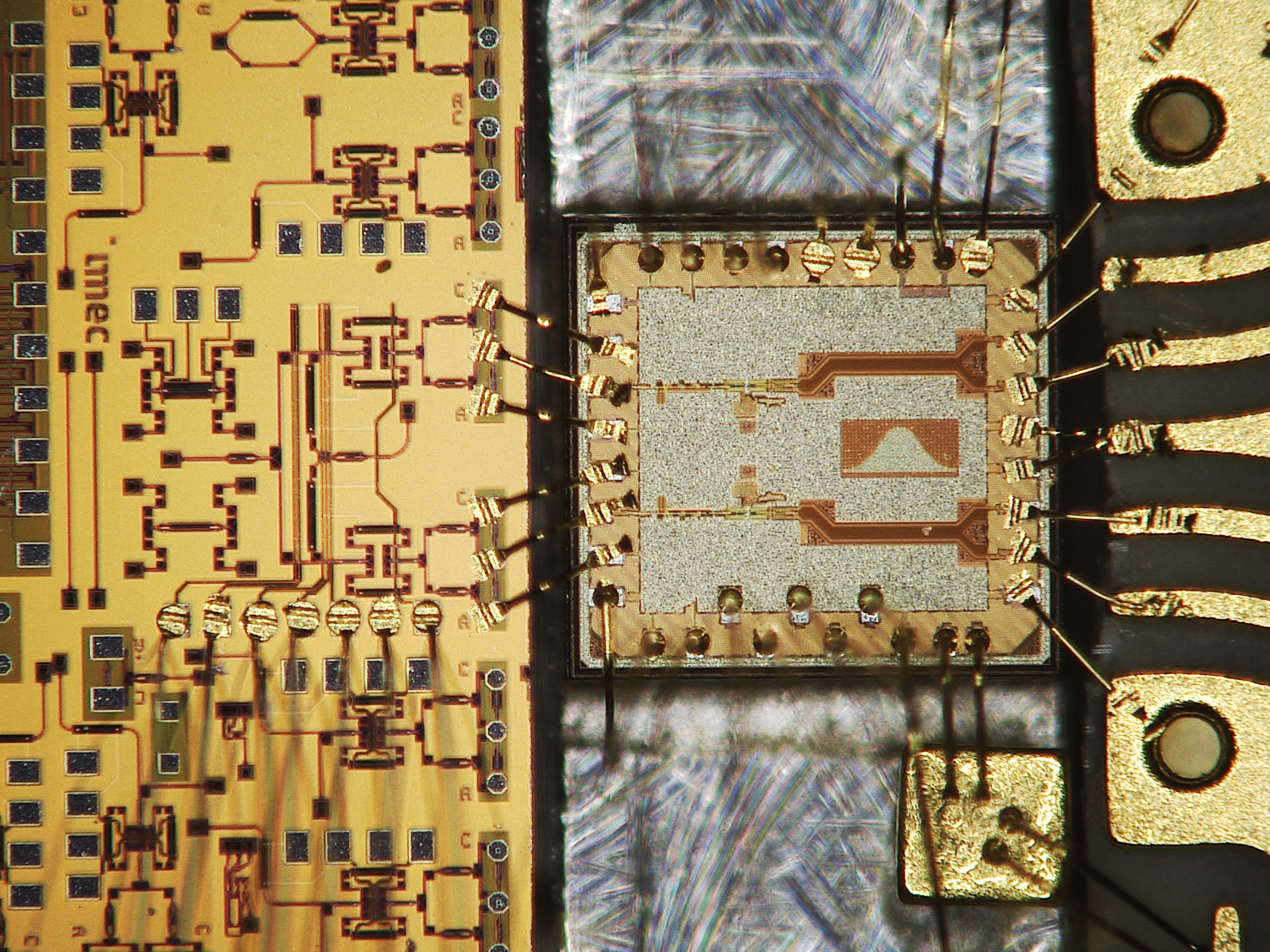}
\caption{Micrograph of the manufactured devices, on the right side is the TIA and on the left side is the PIC.}
\label{assembly}
\end{figure}

This section presents an experimental characterization of the designed balanced homodyne detector. During assembly, the PIC and the TIA are mounted on a heat sink to improve thermal stability and are wirebonded to the custom-designed printed circuit board (PCB). The balanced PDs on the PIC are wirebonded directly to the TIA input. The bond pads of the PIC and the TIA are positioned in close proximity to minimize the bond-wire length and thereby reduce the parasitic inductance. To mitigate the impact of power-supply fluctuations, decoupling capacitors are placed close to the TIA. The complete assembly is shown in Figure \ref{assembly}. The TIA consumes approximately 45 mW of power and occupies a physical area of 1.2 mm x 1.2 mm.

\subsection{S-Parameter Measurement of the TIA}

\begin{figure}[!t]
\centering
\includegraphics[width=3.5in]{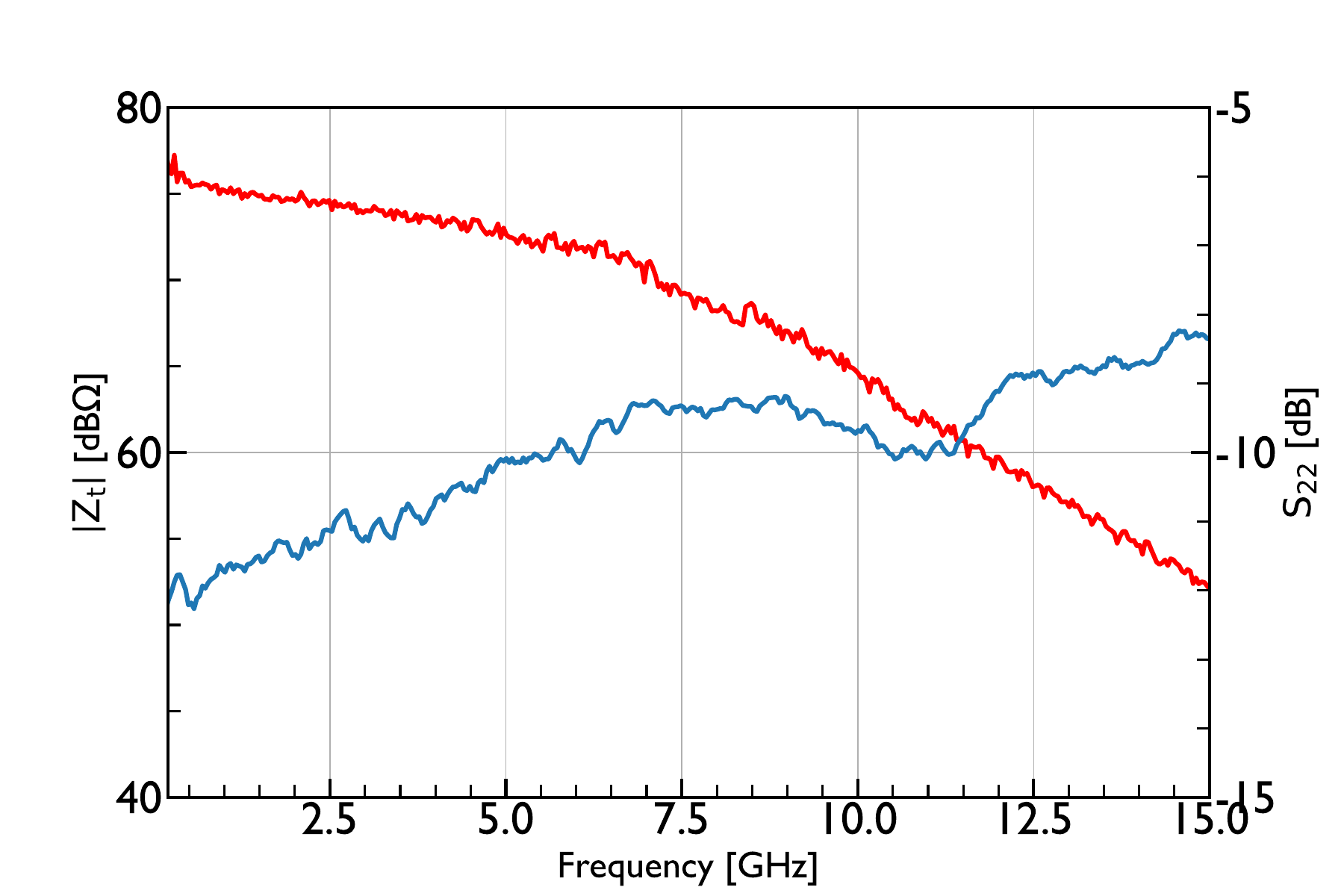}
\caption{Measured transimpedance gain (red) and output matching parameter $S_{22}$ (blue). Output is matched to 50 $\Omega$ when |$S_{22}|$ < -10dB.}
\label{gain}
\end{figure}

The standalone TIA was characterized using S‑parameter measurements. An MPI probing station and an Agilent N5247B PNA‑X network analyzer were used. Figure \ref{gain} shows the measured transimpedance gain and the corresponding $S_{22}$. The transimpedance gain, evaluated at 220 MHz, is equal to 76.2 dB$\Omega$. The 3-dB bandwidth is 3.9 GHz. The output remains matched to 50 $\Omega$ for frequencies where |$S_{22}|$ < -10 dB, which is satisfied up to 5 GHz.   

\subsection{Opto-electrical S-parameter Measurement of the BHD}

\begin{figure}[!t]
\centering
\includegraphics[width=3.8in]{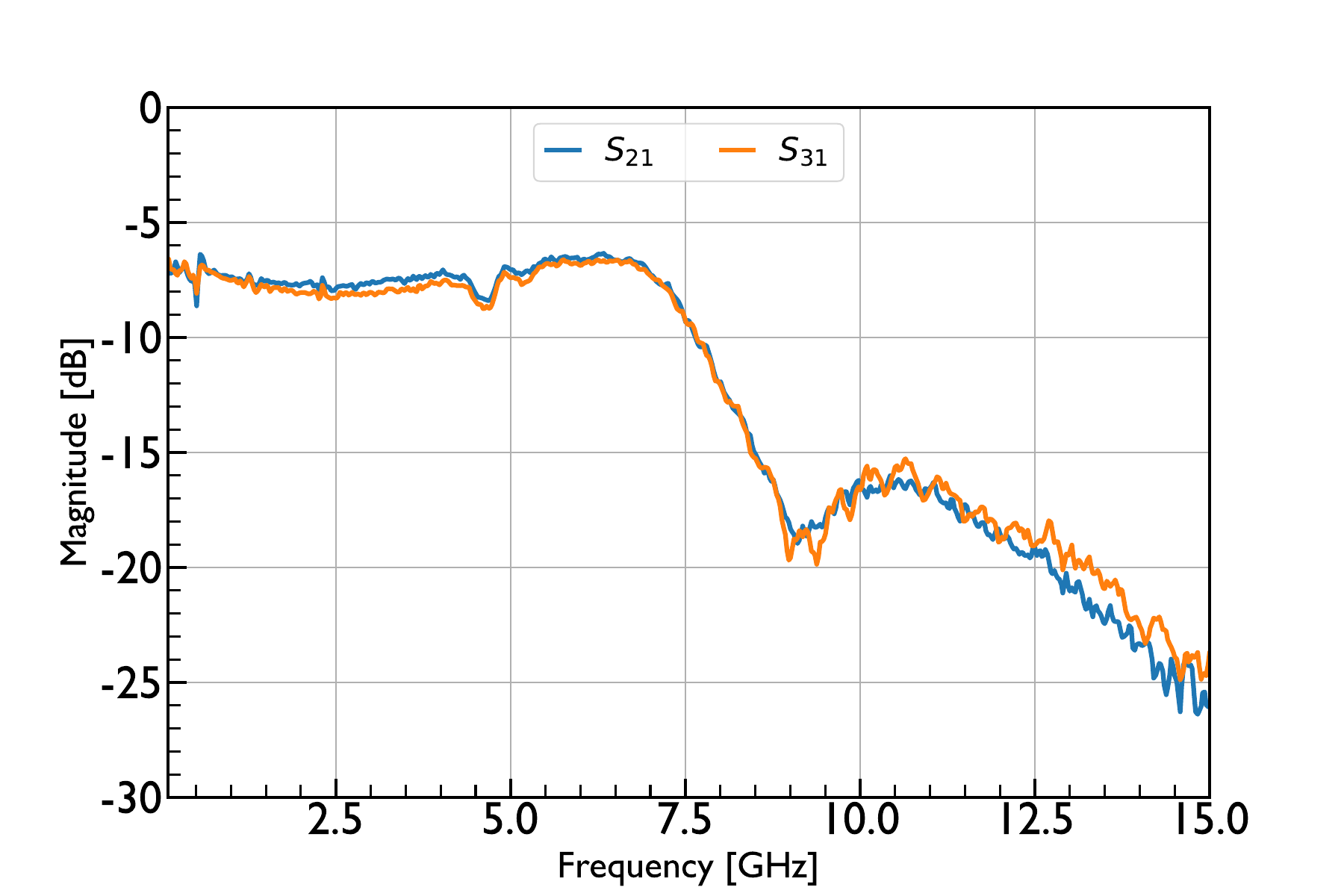}
\caption{Measured opto-electrical $S_{21}$ and $S_{31}$ of the BHD.}
\label{EOs}
\end{figure}

The opto-electrical S-parameters of the BHD were characterized using an external $LiNbO_3$ Mach-Zehnder modulator (MZM) from Fujitsu, with an optical bandwidth exceeding 25 GHz. During the measurement, the current through the PDs was 150 µA. The input port of the MZM and the differential output ports of the TIA were connected to an Agilent N5247B PNA‑X network analyzer. To isolate the response of the BHD, the S‑parameters of the MZM, measured independently using a 70 GHz Finisar XPDV3120 photodetector, were subtracted from the measured $S_{21}$ and $S_{31}$ . Figure \ref{EOs} shows the resulting opto‑electrical $S_{21}$ and $S_{31}$, from which a 3 dB bandwidth of 7.6 GHz is obtained. The two differential output signals exhibit nearly identical responses. The notch arises from parasitic elements, with dominant contributions originating from the PCB traces and coaxial connector. 

\subsection{Noise Measurements of the BHD}
\begin{figure*}[!t]
\centering
\subfloat[]{\includegraphics[width=3.5in]{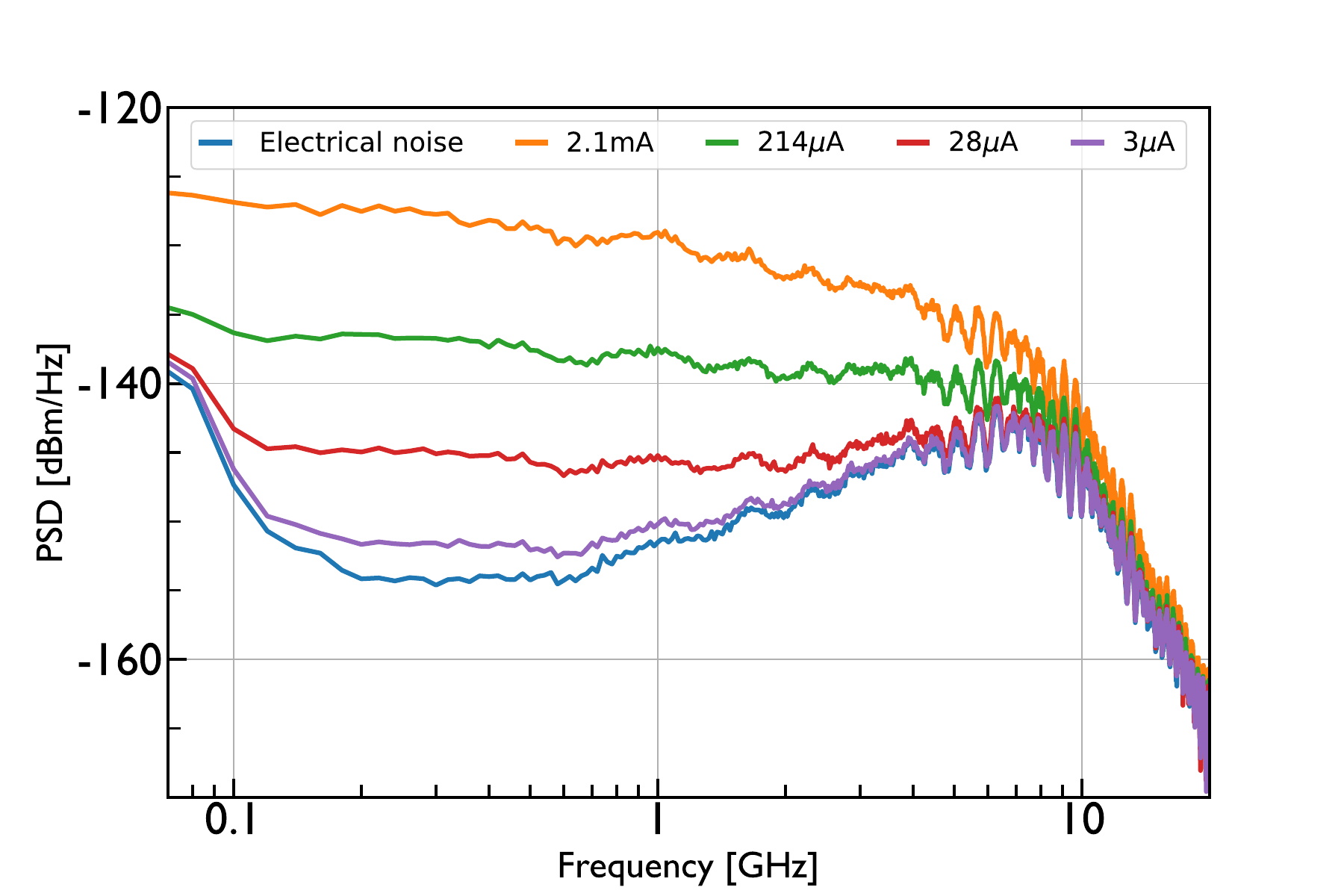}%
\label{noise_psd}}
\hfil
\subfloat[]{\includegraphics[width=3.5in]{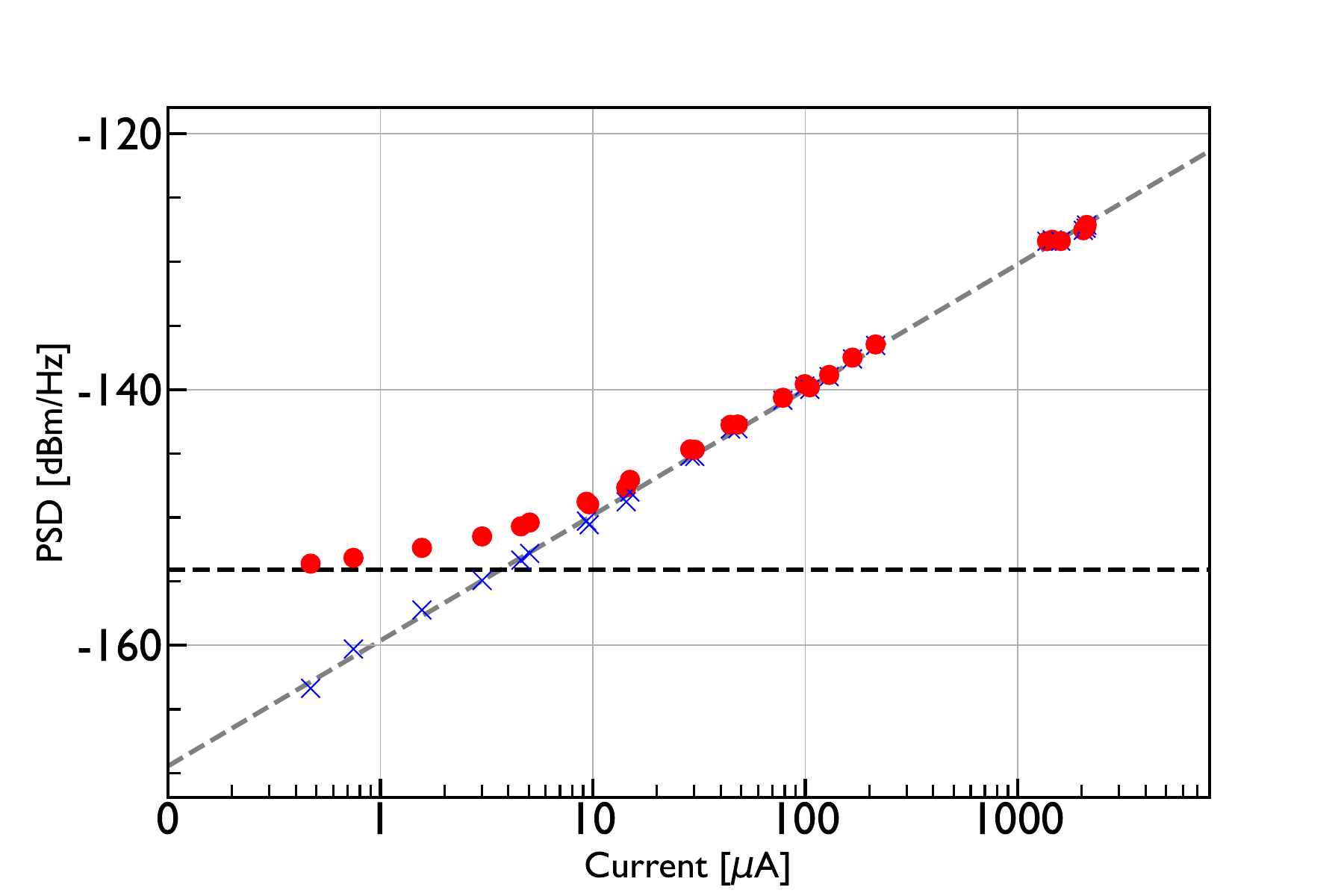}%
\label{noise_psd_freq}}
\caption{(a) Noise PSD at TIA output for different photocurrents. (b) Measured noise PSD at 220 MHz versus photocurrent. The horizontal dashed line marks the measured TIA electronic noise. Red dots correspond to the total measured noise, and blue crosses show the shot noise.}
\end{figure*}

The noise performance of the complete BHD was evaluated by measuring the output noise power spectral density (PSD). A vacuum input was applied to one grating coupler, while the LO was injected into the other. The TIA output was connected to an Agilent N9020A MXA signal analyzer. The LO was generated by a 1550 nm continuous-wave laser (Koheras Basiks E15 source, NKT Photonics), and its power was adjusted using an optical attenuator. A polarization controller (PC) was used to optimize the polarization of the incoming light. Both PDs were reverse-biased by 1 V. The photocurrents were monitored using a Keithley 2450 source meter.

\begin{figure}[!t]
\centering
\includegraphics[width=3.7in]{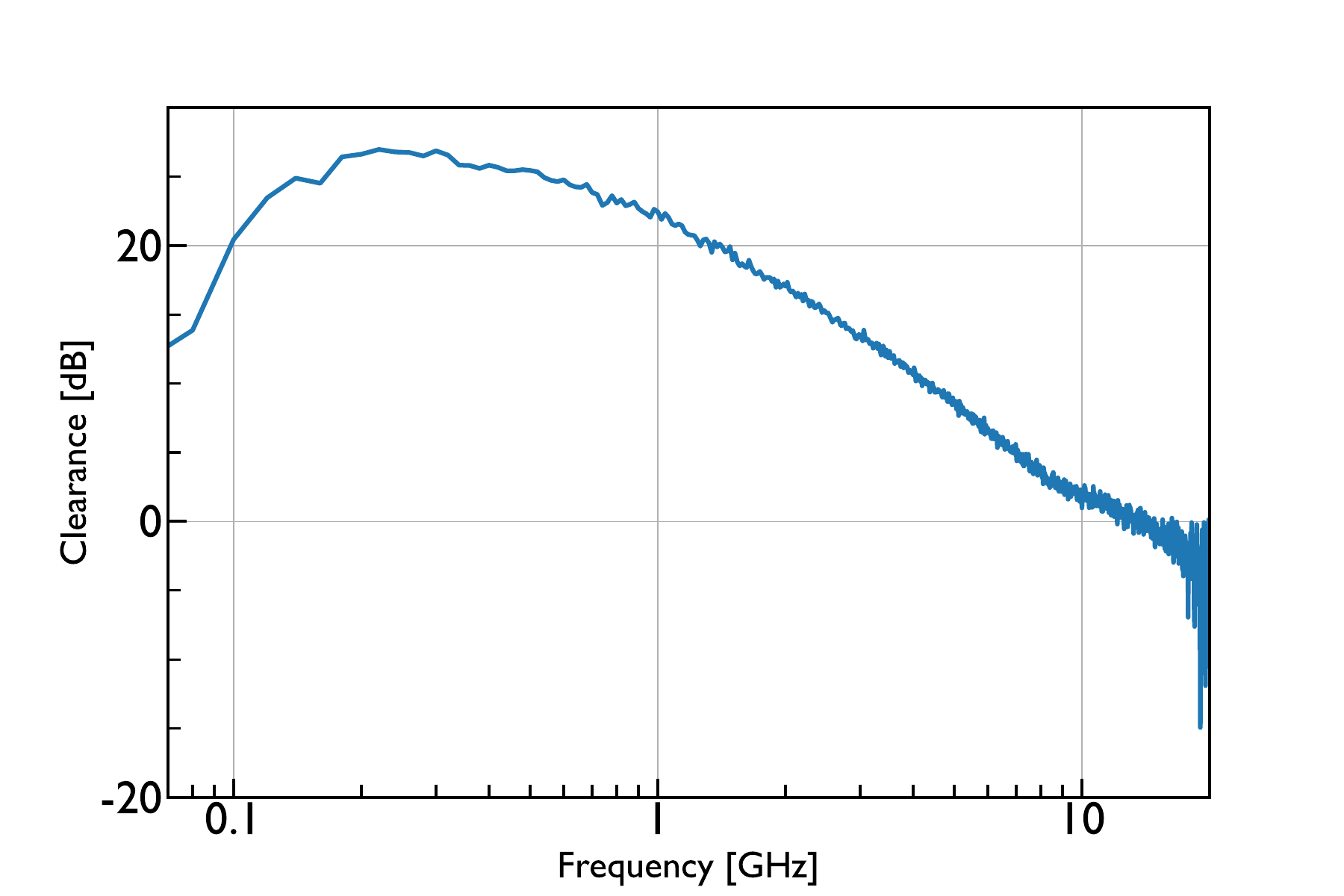}
\caption{Clearance curve measured at 2.2 mA photocurrent.}
\label{clearance}
\end{figure}

Figure \ref{noise_psd} shows the measured output noise PSD as a function of the frequency for different photocurrents. Figure \ref{noise_psd_freq} presents the PSD at 220 MHz as a function of the photocurrent. At low photocurrents, the noise is dominated by the electronic noise from the receiver. As the photocurrent increases, the quantum shot noise becomes the dominant contribution. The clearance, defined as the ratio between the quantum shot noise and the classical noise, is shown in Figure \ref{clearance}. A maximum clearance of 27 dB at 220 MHz is achieved for a photocurrent of 2.2 mA. The shot noise limited bandwidth of the system is 13.7 GHz.

\subsection{Common-Mode Rejection Ratio Measurements}

The common-mode rejection rate (CMRR) quantifies the ability of a BHD to suppress common-mode signals. To measure the CMRR, the laser output was modulated using a sinusoidal signal, and applied to one grating-coupler input. For both balanced and intentionally unbalanced cases, the PDs are reverse-biased at the same voltage. The imbalance for the unbalanced measurement was introduced by adjusting the MZI heater. All measurements were performed using the designed feedback loop. The output spectra were recorded using an Agilent N9020A MXA spectrum analyzer from 10 MHz to 5 GHz. The CMRR was computed using the method described in \cite{cedric}. 

\begin{figure}[!t]
\centering
\includegraphics[width=3.7in]{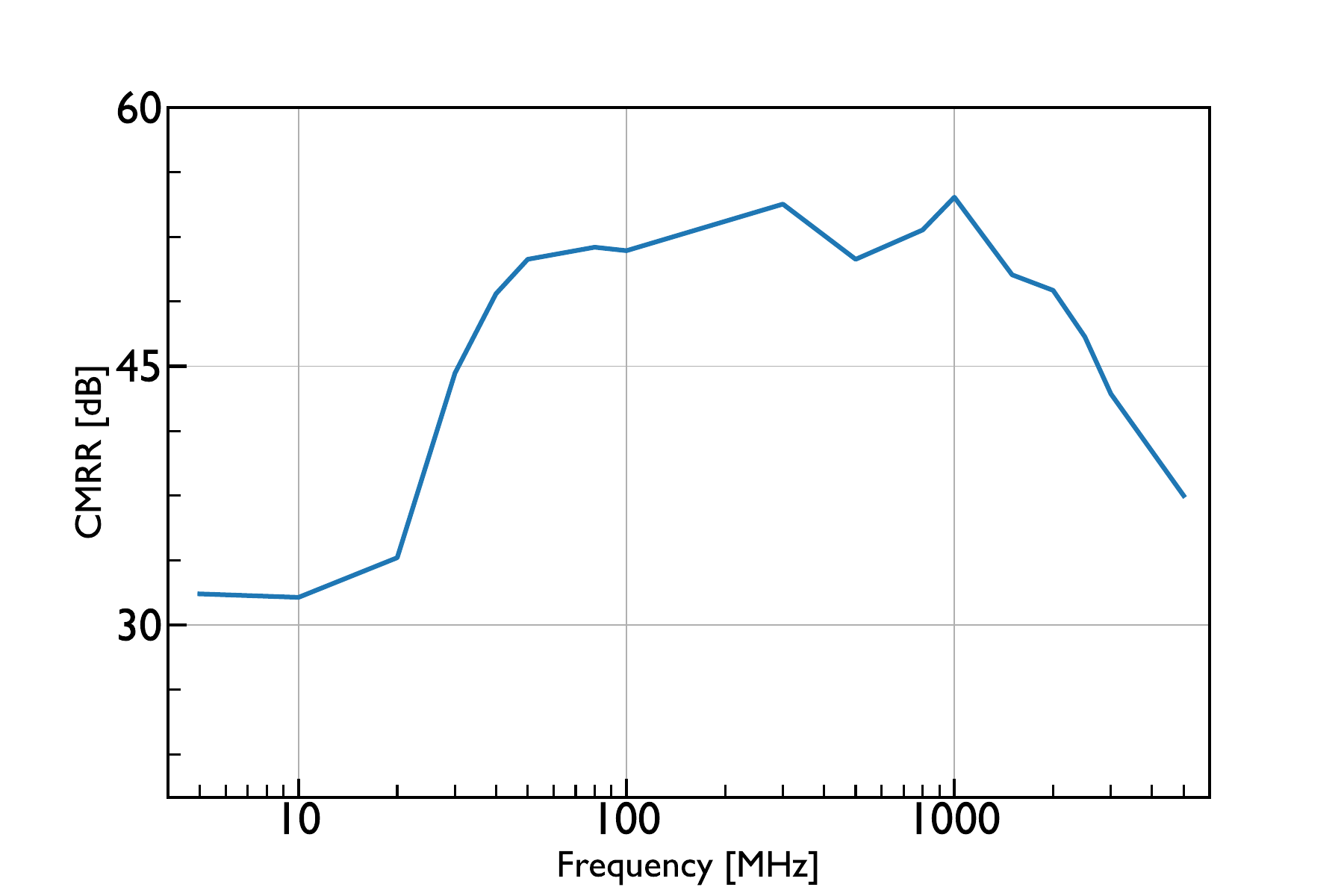}
\caption{Common mode rejection ratio with respect to frequency.}
\label{CMRR}
\end{figure}

The measured CMRR as a function of the frequency is shown in Figure \ref{CMRR}. Across the entire frequency range, the CMRR remains above 31 dB, reaching a maximum value of 54.7 dB at 1 GHz. The reduced CMRR at lower frequencies is attributed to the DC cancellation circuit of the TIA, which suppresses the low-frequency components. A high CMRR demonstrates an effective suppression of the common-mode signals and confirms the proper operation of the analog feedback loop.

\subsection{Comparison to State-of-the-Art Balanced receivers}

\begin{table*}
\tiny
\caption{Comparison to State-of-the-Art BHD}
\resizebox{\textwidth}{!}{\begin{tabular}{c p{0.7cm} p{0.7cm} c c c c}
\hline
\textbf{Reference} & \textbf{BW\textsubscript{3dB}} & \textbf{BW\textsubscript{shot}} & \textbf{Max clearance} & \textbf{CMRR [frequency]} & \textbf{Technology} & \textbf{Power consumption}  \\
This work & 7.6 GHz & 13.7 GHz & 27 dB & 31 dB-54.7 dB[10 MHz-5 GHz] & 28 nm CMOS & 45 mW \\

\cite{bicmos}  &15.3 GHz &26.5 GHz &12 dB &27 dB at 500 MHz & 250 nm BiCMOS & - \\

\cite{cedric} & 1.5 GHz & 20 GHz & 28 dB & 80–26 dB [0.01–20 GHz] &100 nm GaAs pHEMT & 850 mW\\

\cite{APD} & 850 MHz& - &20 dB & 50 dB-30 dB[0.1-1 GHz] & Off-the-shelf & -\\ 

\cite{cbh} & 2.6 GHz & 5 GHz\textsuperscript{1} & 21.1 dB&50 dB[$<$1 GHz] & Off-the-shelf & 56.1 mW \\

\cite{bhdc} & 1.7 GHz& 9 GHz & 14 dB & 52 dB[1 GHz] & Off-the-shelf & 83 mW\textsuperscript{1} \\

\cite{bhd_r} & 4.75 GHz & 23 GHz &12.9 dB & 39.6 dB[1 GHz] & Off-the-shelf & - \\

\cite{C_bhd} & 2.5 GHz & 3.2 GHz\textsuperscript{1} & 15dB & 34.3dB & 180 nm CMOS & 70mW \\

\hline
\end{tabular}}
[1] Values were estimated from the references.
\label{tab1}
\end{table*}

Table \ref{tab1} compares the proposed BHD with state-of-the-art implementations. Despite the inherently poor noise performance of CMOS technologies, the presented design achieves a high clearance comparable to that of leading III–V and discrete implementations, while consuming significantly less power. This highlights the effectiveness of capacitive-feedback topology for low-noise, and high-speed BHD applications. The measured 3-dB bandwidth is also the highest for the non monolithic implementations.

\section{Conclusion}

A capacitive‑feedback transimpedance amplifier is presented, designed using 28 nm CMOS technology, and integrated with a silicon photonic PIC fabricated in imec’s iSiPP200 platform to realize a balanced homodyne detector. The BHD achieves a 7.6 GHz bandwidth with a power consumption of 45 mW, a maximum clearance of 27 dB and a shot‑noise‑limited bandwidth of 13.7 GHz. These results confirm that the proposed architecture is well suited for continuous‑variable quantum key distribution, high‑speed quantum random number generation, and other sensitive optical measurement applications. A CMOS implementation offers low fabrication cost and the potential for co‑integration with digital circuitry.

\end{document}